\documentclass[10pt,journal]{IEEEtran}
\usepackage{times,amssymb,amsmath,amsfonts,float,nicefrac,color,bbm,balance}
\usepackage{euscript,graphics}

\newtheorem{theorem}{Theorem}[section]

\newtheorem{lemma}[theorem]{Lemma}
\newtheorem{proposition}[theorem]{Proposition}
\newtheorem{corollary}[theorem]{Corollary}
\newtheorem{definition}{Definition}[section]

\newcommand{\remove}[1]{}
\renewcommand{\tilde}{\widetilde}

\def\supp{\qopname\relax{no}{supp}}
\def\avg{{\text{\sf E}}}
\def\sgn{\qopname\relax{no}{sgn}}
\def\rank{\qopname\relax{no}{rk}}

\newcommand\nc\newcommand
\nc\bfa{{\boldsymbol a}}\nc\bfA{{\bf A}}\nc\cA{{\mathcal A}}
\nc\bfb{{\boldsymbol b}}\nc\bfB{{\bf B}}\nc\cB{{\mathcal B}}
\nc\bfc{{\boldsymbol c}}\nc\bfC{{\bf C}}\nc\cC{{\mathcal C}}
\nc\bfd{{\boldsymbol d}}\nc\bfD{{\bf D}}\nc\cD{{\mathcal D}}
\nc\bfe{{\boldsymbol e}}\nc\bfE{{\bf E}}\nc\cE{{\mathcal E}}
\nc\bff{{\boldsymbol f}}\nc\bfF{{\bf F}}\nc\cF{{\mathcal F}}
\nc\bfg{{\boldsymbol g}}\nc\bfG{{\bf G}}\nc\cG{{\mathcal G}}
\nc\bfh{{\boldsymbol h}}\nc\bfH{{\bf H}}\nc\cH{{\mathcal H}}
\nc\bfi{{\boldsymbol i}}\nc\bfI{{\bf I}}\nc\cI{{\mathcal I}}
\nc\bfj{{\boldsymbol j}}\nc\bfJ{{\bf J}}\nc\cJ{{\mathcal J}}
\nc\bfk{{\boldsymbol k}}\nc\bfK{{\bf K}}\nc\cK{{\mathcal K}}
\nc\bfl{{\boldsymbol l}}\nc\bfL{{\bf L}}\nc\cL{{\mathcal L}}
\nc\bfm{{\boldsymbol m}}\nc\bfM{{\bf M}}\nc\cM{{\mathcal M}}
\nc\bfn{{\boldsymbol n}}\nc\bfN{{\bf N}}\nc\cN{{\mathcal N}}
\nc\bfo{{\boldsymbol o}}\nc\bfO{{\bf O}}\nc\cO{{\mathcal O}}
\nc\bfp{{\boldsymbol p}}\nc\bfP{{\bf P}}\nc\cP{{\EuScript P}}
\nc\bfq{{\boldsymbol q}}\nc\bfQ{{\bf Q}}\nc\cQ{{\mathcal Q}}
\nc\bfr{{\boldsymbol r}}\nc\bfR{{\bf R}}\nc\cR{{\mathcal R}}
\nc\bfs{{\boldsymbol s}}\nc\bfS{{\bf S}}\nc\cS{{\mathcal S}}
\nc\bft{{\boldsymbol t}}\nc\bfT{{\bf T}}\nc\cT{{\mathcal T}}
\nc\bfu{{\boldsymbol u}}\nc\bfU{{\bf U}}\nc\cU{{\mathcal U}}
\nc\bfv{{\boldsymbol v}}\nc\bfV{{\bf V}}\nc\cV{{\mathcal V}}
\nc\bfw{{\boldsymbol w}}\nc\bfW{{\bf W}}\nc\cW{{\mathcal W}}
\nc\bfx{{\boldsymbol x}}\nc\bfX{{\bf X}}\nc\cX{{\mathcal X}}
\nc\bfy{{\boldsymbol y}}\nc\bfY{{\bf Y}}\nc\cY{{\mathcal Y}}
\nc\bfz{{\boldsymbol z}}\nc\bfZ{{\bf Z}}\nc\cZ{{\mathcal Z}}
\nc\od{{\bar d}}\nc\ow{{\bar w}}\nc\odelta{{\bar\delta}}
\nc\ox{{\bar x}}\nc\oy{{\bar y}}\nc\ou{{\bar u}}
\nc\oh{{\bar h}}

\newcommand\reals{{\mathbb R}}

\newcommand\ff{{\mathbb F}}

\nc\ellone{{\ell_1}}
\nc\elltwo{{\ell_2}}
\nc\ellinf{{{\ell_\infty}}}
\nc\ip[2]{\langle #1,#2\rangle}

\newcommand{\beeq}{\begin{eqnarray*}}
\newcommand{\eneq}{\end{eqnarray*}}

\newcommand{\half}{\nicefrac12}

\nc\red[1]{\textcolor{black} {#1}}

\begin{document}
\sloppy

\title{Restricted isometry property of random subdictionaries}

%\thanks{{\em Date}\/: \today.\/ }%

\author{Alexander Barg~\IEEEmembership{Fellow,~IEEE}, Arya Mazumdar~\IEEEmembership{Member,~IEEE}, Rongrong Wang

\thanks{Manuscript received Jun 4, 2014, revised Feb 9, 2015. }
\thanks{A.~Barg is with the
Dept. of Electrical and Computer Engineering and Institute for Systems
Research, University of Maryland, College Park, MD 20742,
and Institute for Problems of Information Transmission,
Russian Academy of Sciences, Moscow, Russia. Email: abarg@umd.edu.}
  \thanks{A.~Mazumdar is with the Dept. of Electrical and Computer Engineering, University of Minnesota-Twin Cities, Minneapolis, MN 55455. This work was done
partially while the author was at the University of Maryland, College Park, MD. Email: arya@umn.edu.} 
 \thanks{R.~Wang is with the
Dept. of Mathematics, the University of British Columbia, Vancouver, BC, Canada. Email: rongwang@math.ubc.ca.}
  \thanks{This research is supported in part by NSF grants
CCF1217245, CCF1217894, and DMS1101697.}
 \thanks{The results of
  this paper were presented in part at the International Symposium on Information Theory, 2011 \cite{maz11}, and the Allerton conference, 2011 \cite{mb11}.}
}

\maketitle

\begin{abstract}
We study statistical restricted isometry, a property closely related to sparse signal recovery,
of deterministic sensing matrices of size $m \times N$. A matrix is said to have a statistical
restricted isometry property (StRIP) of order $k$ if most   submatrices with $k$ columns define a near-isometric map
of $\reals^k$ into $\reals^m$.
As our main result, we establish sufficient conditions  for the
StRIP property of a matrix in terms of the mutual coherence and \red{mean square coherence.} We show that for many existing deterministic families of sampling matrices,
$m=O(k)$ rows suffice for $k$-StRIP, which is an improvement over the known estimates of either  $m = \Theta(k \log N)$ or $m = \Theta(k\log k)$. \red{We also give examples of matrix families that are shown to have the StRIP property using our sufficient conditions.}
\end{abstract}
\maketitle

\section{Introduction}

\subsection{RIP matrices and binary codes}  We study conditioning properties of 
subdictionaries motivated by the problem of faithful recovery of sparse signals from low-dimensional projections. 
A universal sufficient condition for reliable reconstruction of sparse signals is given by
the restricted isometry property (RIP) of sampling matrices \cite{can08a}. It has been shown 
that sparse high-dimensional signals compressed to low dimension using linear RIP 
maps can be reconstructed using $\ell_1$ minimization procedures such as Basis pursuit
and Lasso \cite{can05,can06b,can08a,cai09}.

Let $\bfx$ be an $N$-dimensional signal and denote by $[N]=\{1,2,\dots, N\}$ the set of coordinates. Below we use $\Phi$ to denote the $m\times N$ sampling matrix and write $\Phi_I$ to refer to the $m\times k$ submatrix of $\Phi$
formed of the columns with indices in $I$, where $I=\{i_1,\dots,i_k\}\subset [N]$ is a $k$-subset of $[N].$
We say $\Phi$ is $(k,\delta)$-RIP if every $k$ columns of $\Phi$ satisfy the following near-isometry property:
\begin{equation}\label{eq:quasi}
\|\Phi_I^T\Phi_I-\text{Id}\|_2\leq \delta 
\end{equation}
where $\text{Id}$ is the identity matrix,
and $\|\cdot\|_2$ is the spectral norm (the largest singular value).

\red{It is known that a $k$-RIP matrix must have at least $m=\Omega(k\log(N/k))$ rows
\cite{kas77, GG84}. Moreover, if $\bfx$ is compressed to a sketch $\bfy=\Phi \bfx$ of dimension $m$, then
$m=\Omega(k\log(N/k))$ samples are required for any recovery algorithm to provide an approximation of the signal with an error
guarantee expressed in terms of the $\ell_1$ or $\ell_2$ norm \cite{kas07,kha10} (this bound applies to signals which are
not necessarily $k$-sparse). Matrices with random Gaussian or Bernoulli entries with high probability provide 
the best known error guarantees for recovery from sketches of dimension $m$ that matches this lower bound \cite{can05,can06a,can06c}.}

\remove{We recall the well-known parameters, the mutual coherence $$\mu:=\max_{i \ne j}\{\langle \phi_i,\phi_j \rangle\},$$
 and the matrix operator norm above, used in \cite{Step12,tro08b,tro08}. }

Let $\mu_{i,j}=|\langle\phi_i,\phi_j\rangle|$ be the \red{ coherence} between columns $i$ and $j$
and denote by $\mu:=\max_{i \ne j}\mu_{i,j}$ the {\em \red{mutual coherence parameter}} of the matrix $\Phi.$
The relation between the mutual coherence and RIP has served the
starting point in a number of studies on RIP matrix construction \cite{tro04},
\cite{don03}. 
\red{One way of constructing incoherent dictionaries begins with taking a binary code, i.e., a
set $\cC$ of binary $m$-dimensional vectors. We say that the code $\cC$ has small width if all
pairwise Hamming distances between distinct vectors of $\cC$ are close to $m/2$. For instance, if
$m/2-w\le d(x_i,x_j)\le m/2+w$ for every $x_i,x_j\in \cC, x_i\ne x_j,$ we say that the code has width $w$.}
\red{A real sampling matrix can be generated from a small-width binary code by mapping bits of the codewords
to bipolar signals according to $0\to 1, 1\to -1.$  The resulting vectors are normalized to unit length and written
 in the columns of the matrix $\Phi.$}
The \red{coherence parameter} $\mu(\Phi)$ \red{of the matrix} and the width \red{of the code} $\cC$ are connected by 
\red{the obvious equality} $w(\cC)=\mu(\Phi) m/2.$

One of the first papers to put forward the idea of constructing
RIP matrices from binary vectors was \cite{dev07}.
While it did not make a connection to error-correcting codes,
a number of later papers pursued both its algorithmic and constructive
aspects \cite{bar10,cal10a,cal10b,dai09}.
Examples of codes with small width are given in \cite{alo92}, where they are
studied under the name of small-bias probability spaces.
RIP matrices obtained from the constructions in \cite{alo92} satisfy
$m=O(\frac{k\log N}{\log(\log k N)})^2$.
In \cite{ben09} these results were
recently improved to $m=O(\frac{k\log N}
{\log k})^{5/4}$ for $(\log N)^{-3/2}\le \mu\le (\log N)^{-1/2}.$ 
The advantage of obtaining RIP matrices from binary or spherical codes
is low construction complexity: in many instances it is possible to define
the matrix using only $O(\log N)$ columns while the remaining columns can
be computed as their linear combinations.
We also note a result of \cite{bou10} that
gave the first (and the only known) construction of RIP matrices with $k$ on the
order of $m^{\frac 12+\epsilon}$
(i.e., greater than $O(\sqrt m)$). An overview of the
state of the art in the construction of RIP matrices is given in
a recent paper \cite{bandeira12}. 

 Taking the point of view that constructions
of complexity $O(N)$ are acceptable,
the best tradeoff between $m,k$ and $N$ for RIP-matrices based on codes and \red{mutual} coherence is obtained
from Gilbert-Varshamov-type code constructions \cite{por08}: namely, it is possible
to construct $(k,\delta)$-RIP matrices with $m=4(k/\delta)^2\log N$.
At the same time, already the results of \cite{alo92} imply that
the sketch dimension in RIP matrices constructed from binary codes is
at least $m=\Theta((k^2\log N)/\log k).$

\subsection{Statistical RIP (StRIP) matrices}
Constructing deterministic RIP matrices or verifying that a matrix satisfies the RIP is a difficult problem. For this reason
in order to approach the optimal sketch dimension  $O(k\log N/k)$ we focus on the following probabilistic relaxation of 
definition \eqref{eq:quasi}.
\begin{definition}[Statistical Restricted Isometry Property]
Let $\Phi$ be an $m\times N$ real matrix, where $m\leq N.$ Suppose that $I\subset N, |I|=k$
is chosen uniformly at random from $[N].$  Then $\Phi$ is said to have the $(k,\delta, \epsilon)$-StRIP if 
\begin{equation*}
P(\|\Phi_I^T\Phi_I-\text{\rm Id}\|_2\geq \delta)<\epsilon .
\end{equation*}
\end{definition}
Except for the name, the StRIP is by no means new in the literature. Tropp \cite{tro08} showed how StRIP and a condition on the so called local 2-cumulative coherence
\[
%\mu_2(T)=\max\big[\sum\limits_{j\in T}|\langle \phi_j,\phi_k\rangle |^2 \big]^{1/2}
\mu_2(T)=\max\limits_k \big[\sum\limits_{j\in T}\mu_{j,k}^2 \big]^{1/2}
\]
can support sparse recovery of a class of signals. Cand{\`e}s and Plan \cite{can09a} used the same technique to prove almost exact 
recovery for the Lasso estimator. 

StRIP is a property of interest in its own right, apart from applications in sparse recovery. Indeed, papers such as
\cite{tro08} are entirely devoted to bounds on the largest singular value of a random collection of columns from a general dictionary. 
The recent paper \cite{Step12} states that
StRIP is ``of great potential interest for a wide class of problems involving
high-dimensional linear or nonlinear regression models.'' \cite{Step12} goes on to investigate sufficient conditions
for StRIP based on the mutual coherence of the matrix $\Phi.$

 The goal of this paper is to broaden the class of StRIP matrices by establishing a sufficient condition 
 that relies upon easy-to-verify parameters of sampling matrices. 
  In this vein, we introduce a new parameter called the {\em mean square coherence}
   \[
\bar \mu^2=\max_{1\le j\le N}\frac1{N-1} \sum_{\substack{i=1\\i\ne j}}^N\mu_{i,j}^2.
   \]
In many cases, as we will see below, calculations with the mutual coherence parameter can  be too pessimistic.
In this paper we combine the mean square and mutual coherence parameters to relax the requirements on camping matrices. 
%We note that A somewhat similar average coherence condition was also
%introduced in \cite{baj10a,baj11}.

Intuitively, the mean square coherence parameter is easier to control than $\mu(\Phi).$ 
%can be more suitable than the mutual coherence when we only care about the behavior of most of the subdictionaries.
Note that if the matrix $\Phi$ is {\em coherence-invariant} (i.e., the set $M_i:=\{\mu_{ij}, j\in [N]\backslash i\}$
is independent of $i$), then $\bar \mu^2$ can be computed for any given $\phi_j$ without finding the maximum. Observe that most known constructions of sampling matrices satisfy this property. This includes matrices constructed from linear codes \cite{dev07,bar10},
chirp matrices and various Reed-Muller matrices \cite{baj10a,cal10a}, as well as subsampled Fourier matrices \cite{hau10}.

The main contribution of this paper is the derivation of  new sufficient conditions for the StRIP property of sampling matrices, stated in Theorem \ref{thm:main}. The proof of this theorem is based on 
considering the {mean square coherence} $\bar{\mu}^2$ and on detailed analysis of statistical incoherence of sampling matrices. The sufficient conditions that arise are 1) phrased in terms of coherence $\mu$ and $\bar{\mu}^2$,  2) easy to verify and 3) analytically easy to evaluate for many known families of sampling matrices. 
We show that our results are better than the estimates
known in the literature for a range of  the sparsity and the signal dimension that  satisfy conditions discussed in Sec.~\ref{sec:comp}.
%which in practice is a rather large region of the parameters.
 In general, Theorem \ref{thm:main} extends
the currently known region of sufficient conditions for StRIP matrices, and for many standard sampling matrices, ensures that $m=O(k)$ rows suffice for $k$-StRIP, which is an improvement over the known estimates of   $m = \Theta(k \log N)$.

Application of our results to some deterministic
matrices popularized in recent literature on sparse recovery, for instance, the Delsarte-Goethals matrices \cite{cal10a,cal10b}, shows that the statistical RIP property is fulfilled for a smaller sketch 
dimension $m$ than previously known. We also estimate the dimensions of many other known families of matrices,
deriving sufficient conditions for the statistical RIP property. Since the StRIP and statistical incoherence properties suffice for stable recovery with Basis Pursuit, our results, in turn,
provide sufficient conditions for sparse recovery for many families of sampling matrices. A more detailed discussion and some further applications of our results 
appear in an earlier version of this paper in 
arXiv \cite{bmw2013}.

\section{Main result and discussion}
\subsection{Main result}
%\begin{theorem}\label{thm:main} Let $\Phi$ be an $m\times N$ matrix with unit-norm columns, mutual coherence $\mu$
%and mean square coherence $\bar{\mu}^2.$ Suppose we have
%\[
%\mu\leq c_1m^{-1/4},\quad \bar{\mu}^2 \leq c_2 m^{-1/2}, \quad  \|\Phi\|^2_2\leq c_3\frac{N}{k} 
%\]
%then $\Phi$ is $(k,\delta, \epsilon)$-StRIP as long as $m\geq c_4k \log k \log^3 1/\epsilon$. Here,  $c_1$, $c_2$ are pure constants, $c_3$ is a function of $\delta
%$, and $c_4$ is a function of  $c_1$, $c_2$, $c_3$, and $\delta$.
%\end{theorem}

\begin{theorem}\label{thm:main} Let $\Phi$ be an $m\times N$ matrix.
Let $\epsilon<\min\{1/k,e^{1-1/\log 2}\}$ and suppose that $\Phi$ satisfies
  \begin{align}\label{eq:sc}
  k\mu^4\le &\frac 1{\log^2(1/\epsilon)}\min\Big(\frac{(1-a)^2b^2}{32\log(2k)\log(e/\epsilon)},{c^2}\Big)\\
  \quad
\text{and}\quad& k\bar{\mu}^2\le\frac{ab}{\log(1/\epsilon)},
       \end{align}
where $a,b,c\in(0,1)$ are constants such that
  \begin{equation}\label{eq:abc}
  \sqrt{a}+\sqrt{2ab}+\sqrt c+\frac {2k}N\|\Phi\|^2\le e^{-1/4}\delta/{6\sqrt 2}.
  \end{equation}
Then $\Phi$ is $(k,\delta,\epsilon)$-StRIP.
\end{theorem}

\subsection{Comparison to earlier work}\label{sec:comp}
Most relevant to our results are two papers by Tropp \cite{tro08b}, \cite{tro08}. The first of them proved a nearly optimal sufficient condition for StRIP using mutual coherence and matrix norm, namely that $\Phi$ is $(k,\delta,\epsilon)$-StRIP if 
\begin{equation}\label{eq:re1}
\mu = O((\log N)^{-1}) \;\text{ and }\;\|\Phi\|^2 = O\left(\frac{N}{k\log N}\right).
\end{equation}
where the constants that depend on $\delta$ are absorbed into $O(\cdot)$. 
For  the above result to hold, $\epsilon$ has to be less than 
$1/k$, just as in Thm.~\ref{thm:main} above. 
The restriction on $\mu$ is very mild, while the 
condition on $\|\Phi\|$ can be further improved. Namely, \cite{tro08} shows that the conditions
\begin{equation}\label{eq:re2}
\mu = O((k\log k)^{-1/2}) \text{ and }\|\Phi\|^2 = O\left(\frac{N}{k}\right)
\end{equation}
suffice for the $(k,\delta, \epsilon)$-StRIP property.  %In both the above results  $\epsilon$ can be $\leq 1/k$.
Note that the improvement for $\|\Phi\|$ in \eqref{eq:re2} over \eqref{eq:re1} is obtained at the expense of tightening the condition
 on the coherence. For this reason, conditions \eqref{eq:re1} are better suited for verifying the StRIP property of deterministic matrices. 

Equations \eqref{eq:re1} and \eqref{eq:re2} together define the currently known region of sufficient conditions for StRIP matrices. The contribution of Theorem \ref{thm:main} is to further extend this 
region by including matrices that satisfy  
\begin{equation}\label{eq:3}
\mu=O( (k\log k)^{-1/4}), \,\bar\mu^2= O(1/k)\text{ and }\;\|\Phi\|^2 = O\left(\frac{N}{k}\right).
\end{equation}

We can claim an improvement over the results of \cite{tro08b} when inequality \eqref{eq:3} is better than \eqref{eq:re1} (in the sense that a smaller value of $m$ is required for the  conditions to be satisfied). 
Most known examples of deterministic sampling matrices, including the examples in Sect. \ref{sec:determin}
below, have mean square coherence of order $\bar \mu^2(\Phi)=O(\frac{1}{m})$, coherence $\mu=\frac{1}{\sqrt m}$  and spectral norm $\|\Phi\|^2\leq \frac{N}{m}$. Hence the most restrictive constraint of the three conditions in \eqref{eq:3} 
is the last one, and  \eqref{eq:3} essentially reduces to the constraint $m = \Theta(k)$ for many standard sampling matrix families. 
On the other hand, \eqref{eq:re1} reduces to the constraint $m =\Theta(k \log N)$ for the same reason. Note that the most restrictive condition in \eqref{eq:re2} is the first one which gives rise to the constraint $m = \Theta(k \log k)$ for the sampling matrices of Sect. \ref{sec:determin}.

The sufficient condition on the coherence $\mu$ implied by \eqref{eq:3}  is 
\begin{equation}\label{eq:re3}
\mu=O( (k\log k)^{-1/4}),
\end{equation}
which by itself is  an improvement over 
the coherence condition of \eqref{eq:re1} 
if $k\log k= O( \log^4 N)$. %, a rather large range in a practical scenario. 
In the next subsection we discuss a concrete family of sampling
matrices for which our results yield better parameters than the conditions known previously.

Apart from this, we also note that imposing the StRIP condition together with the statistical incoherence condition, or SINC (defined below), suffices to prove stable sparse recovery by Basis Pursuit. 
This observation, which is an extension of  known results, is included in the Appendix. We list examples of dictionaries that meet the StRIP and SINC conditions in Sect.~\ref{sec:determin}.
\subsection{Example: Delsarte-Goethals codes}\label{sec:dels}
A class of sensing matrices that satisfy the condition of Theorem \ref{thm:main} comes from a family of binary codes called
the Delsarte-Goethals codes which are  certain nonlinear subcodes of the second-order Reed-Muller codes; see \cite{mac91}, Ch.~15.
Suppose that the length of the chosen code is $m$.
Writing the code vectors as columns of the matrix and replacing $0$ with $1/\sqrt{m}$ and $1$ with $-1/\sqrt{m}$, we obtain
the following parameters:
  \begin{equation}\label{eq:example}
  m=2^{2s+2}, \;N=2^{-r}m^{r+2},\;\mu=2^rm^{-\half}
  \end{equation}
where $s\ge 0$ is any integer, and where for a fixed $s$, the parameter $r$ can be any number
in $\{0,1,\dots,s-1\}.$
If we take $s$ to be such that $s+1$ is divisible by $3$ and set $r=(s+1)/3$, then we obtain,
   $$
m= 2^{6r},\; N=2^{6r^2+11r},\; \mu=2^{-2r} = m^{-1/3}.
   $$
An easy calculation that relies on the Pless identities for binary codes (e.g. \cite[p.132]{mac91}) shows that
%consider Lemma \ref{lem:pless} below.
%\begin{lemma} {\rm (Pless identities, e.g. \cite[p.132]{mac91})}
%\label{lem:pless}
%Let $\cC$ be an orthogonal array of strength $t$.  Let $B_w=(1/N)|\{
%(\phi_i,\phi_j)\in\cC^2\mid d_{ij}=w\}|$ be the number of pairs vectors in
%$\cC$ at distance $w$.  For all $l=1,2,\dots,t$
%   \begin{equation}\label{eq:ppm}
%  \sum_{w=0}^m \frac{B_w}{N}\Big(w-\frac m2\Big)^l=\frac1{2^m}\sum_{w=0}^m\binom mw
%       \Big(w-\frac m2\Big)^l.
%  \end{equation}
%  where $d_{ij}=\frac m2(1-\phi_i^T\phi_j)$ be the Hamming distance between $\phi_i$ and $\phi_j.$
%\end{lemma}
%This lemma  implies that
   \begin{equation}\label{eq:mu21}
\bar\mu^2=\frac{N-m}{m(N-1)}<\frac{1}{m}.
   \end{equation}
Using the properties of the Delsarte-Goethals codes, it is easy to see that the norm of the sampling matrix $\Phi$ is $\|\Phi\|=\sqrt{N/m}$. Employing condition \eqref{eq:re3}, we observe that $m = O(k \log k)$ samples suffice for this matrix to satisfy the $(k,\delta,1/k)$-StRIP condition while \eqref{eq:re1} requires $m=O(k\log N)$.  \red{If $m$ is fixed as above, this implies that using our results we can claim the StRIP property
for larger $k$ that was previously known. }
\remove{\textcolor{blue}{Moreover, we can explicitly calculate that for a fixed $r$, how does $k$ compare with $\log^4 N$. Let $r=2$, then $m=2^{12}$, $N=2^{48}$, hence $\frac{m}{\log^4N}=7.716\times 10^{-4}$. Therefore, as long as $k\leq m$, it is less than $\log^4 N$.[RW: this kind of argument seems good, but it requires us to explicitly find the constant embedded in $O(\cdot)$ in both \eqref{eq:re1} and \eqref{eq:re3} for a fair comparison. If you feel this is not necessary, we can just save the effort and delete this paragraph.}}
\section{Proof of the main result}
\subsection{Notation}\label{sec:notation}
Let $\Phi$ be denote the $m\times N$ real sensing matrix with columns of unit norm. 
By $\cP_k(N)$ we denote the set of all $k$-subsets of $[N].$ 
%Below
%we write $\Phi_I$ to refer to the $m\times k$ submatrix of $\Phi$
%formed of the columns with indices in $I$. Given a vector
%$\bfx\in \reals^N,$ we denote by $\bfx_I$ a $k$-dimensional vector
%given by the projection of the vector $\bfx$ on the coordinates
%in $I$.  
The usual notation 
for probability $\Pr$ is used to refer a probability measure when there is no ambiguity. At the same time, we use
separate notation for some frequently encountered probability spaces.
In particular, we use $P_{R_k}$ to denote the uniform probability distribution on $\cP_k(N)$. 
We also use $P_{R_k'}$ to denote the uniform distribution on the set $R_k':=\{(I,j): |I|=k, I \subseteq [N], j\in I^c\}$. 
%If we need to choose a random $k$-subset $I$ and a random index in $[N]\backslash I,$  

To express our results concisely we introduce the following concept.
\begin{definition} \label{def:sinc}
An $m\times N$ matrix $\Phi$ is said to satisfy a statistical incoherence condition (is $(k,\alpha,\epsilon)$-SINC)
if 
  \begin{equation}\label{eq:sinc}
    P_{R_k}(\{I\in \textstyle{\cP_k(N)}: \max_{i\not\in I}\|\Phi_I^T\phi_i\|_2^2\le\alpha\})\ge 1-\epsilon.
  \end{equation}
\end{definition}
This condition is discussed in \cite{fuc04,tro05}, and more explicitly in \cite{tro08b}.
Following \cite{tro08b}, it appears in the proofs of sparse
recovery in \cite{can09a} and below in this paper. 

  The reason that \eqref{eq:sinc}
is less restrictive than the constraint on the coherence parameter $\mu(\Phi)$ is as follows.
The columns of $\Phi$ can be considered as points in the real projective space
$\reals P^{m-1}.$ 
Recall that $\mu(\Phi)=\min_{i\ne j} |\langle\phi_i,\phi_j\rangle|.$
The columns of a matrix $\Phi$ with small $\mu(\Phi)$ form a packing of the
space with large pairwise separation between the points.
Such a packing cannot contain too many elements 
so as not to contradict universal bounds on packings of $\reals P^{m-1}.$ At the same time, for the norm
$\|\Phi_I^T\phi_i\|_2$ to be large it is necessary that a given column is close to the majority of the $k$ vectors from the set $I$, which is easier to rule out.

%%%%%%%%
% CORRELATION INVARIANT MATRICES
%%%%%%%%
\subsection{Sufficient conditions for statistical incoherence properties}
We begin with establishing a sufficient condition for the SINC property in terms
of the coherence parameters of $\Phi.$ This result is not necessarily stronger than the result of \cite{tro08b}, but is essential in proving our main theorem.
%Let $\bar\mu^2(\Phi):=\avg_{i,j}\mu_{ij}^2$ be the mean square coherence of the matrix $\Phi.$
%

\begin{theorem}\label{thm:sinc} Let $\Phi$ be an $m\times N$ matrix with unit-norm columns, coherence $\mu$
and mean square coherence $\bar{\mu}^2.$ 
  \begin{equation}\label{eq:mu2a}
  \mu^4\le \frac{(1-a)^2\beta^2}{32 k(\log 2N/\epsilon)^{3}}\text{\quad and \quad} \bar{\mu}^2\le \frac{a\beta}{k\log(2N/\epsilon)} ,
  \end{equation}
  where $\beta>0$ and $0<a<1$ are any constants.
  Then $\Phi$ has the $(k,\alpha,\epsilon)$-SINC property with $\alpha=\beta/\log(2N/\epsilon).$
  
%  If $\Phi$ is not coherence-invariant, then the conclusion of the theorem holds if
%$\bar\mu^2$ in \eqref{eq:mu2a} is replaced with $\bar\mu_{\max}^2.$
\end{theorem}

Before proving this theorem we will introduce some notation. Fix $j\in[N]$ and let $I_j=\{i_1,i_2,\dots,i_k\}$
be a random $k$-subset such that $j\not\in I_j.$ The subsets $I_j$ are chosen from the set \red{$[N]\backslash j$} with
uniform distribution. Define random variables $Y_{j,l}=\mu^2_{j,i_l}, l=1,\dots,k.$ Next define
a sequence of random variables $Z_{j,t},t=0,1,\dots,k,$ where
  \begin{align*}
   Z_{j,0}&=\avg_{I_j}\sum_{l=1}^kY_{j,l},\\\quad Z_{j,t}&=\avg_{I_j} \Big(\sum_{l=1}^k Y_{j,l}\mid Y_{j,1},Y_{j,2},
   \dots,Y_{j,t}\Big), \;t=1,2,\dots,k.
 \end{align*}
%From the assumption of coherence invariance, the variables $Z_{j,t}$ for different $j$ are stochastically equivalent.
For $t=1,\dots,k,$ let
  $$
  Z_t=\avg_j Z_{j,t}=\avg_{R_k'}\Big(\sum_{l=1}^k Y_{j,l}\mid Y_{j,1},Y_{j,2},
   \dots,Y_{j,t}\Big), 
  $$
  \red{where $R_k'$ is defined in Section \ref{sec:notation}}.
  
%The random variables $Z_t$ are defined on the set of $(k+1)$-subsets of $[N]$ with probability distribution $P_{R_k'}$ (\red{RW: this sentence does seem weird, maybe delete it?}). We will show
\red{Let us show that the random variables $Z_t$
form a Doob martingale}. Begin with defining a sequence of $\sigma$-algebras $\cF_t,t=0,1,\dots,k,$
where $\cF_0=\{\emptyset,[N]\}$ and $\cF_t, t\ge 1$ is the smallest $\sigma$-algebra with respect to
which the variables $Y_{j,1},\dots,Y_{j,t}$ are measurable (thus, $\cF_t$ is formed of all subsets of $[N]$ of size $\le t+1$).
Clearly, $\cF_0\subset\cF_1\subset\dots\subset\cF_k$, and for each $t,$ $Z_t$ is a bounded random variable that
is measurable with respect to $\cF_t.$ Observe that 
  \begin{align}
  Z_0&=\avg_j Z_{j,0}=\avg_{R_k'}\sum_{l=1}^k \mu_{j,i_l}^2=\sum_{l=1}^k\avg_{R_k'} \mu_{j,i_l}^2\le k\bar\mu^2.
  \label{eq:ce}
    \end{align}
%  \\
%  &\le k\bar\mu^2, \label{eq:nce}

%where \eqref{eq:ce} assumes coherence invariance, and \eqref{eq:nce} is valid independently
%of that assumption.

The next two lemmas are useful in proving Theorem \ref{thm:sinc}.
\begin{lemma}\label{lemma:bounded}
The sequence $(Z_t,\cF_t)_{t=0,1,\dots,k}\;$ forms a bounded-differences martingale, namely
  $
  \avg_{R_k'}(Z_t\mid Z_0,Z_1,\dots,Z_{t-1})=Z_{t-1}$
  and
  $$
  |Z_t-Z_{t-1}|\le 2\mu^2\Big(1+\frac k{N-k-2}\Big), \quad t=1,\dots,k.
  $$
\end{lemma}
\begin{IEEEproof} In the proof we write $\avg$ instead of $\avg_{R_k'}.$ We have
       \begin{align*}
Z_t &= \avg \Big( \sum_{l=1}^{k} Y_{j,l} \mid \cF_t \Big)
=  \sum_{l=1}^{t} Y_{j,l} + \avg \Big( \sum_{l=t+1}^{k} Y_{j,l} \mid \cF_t \Big)\\
&=  Z_{t-1} +Y_{j,t} + \avg \Big( \sum_{l=t+1}^{k} Y_{j,l} \mid \cF_t \Big) - \avg \Big( \sum_{l=t}^{k} Y_{j,l} \mid \cF_{t-1} \Big).
\end{align*}
Next,
  \begin{align*}
\avg (Z_t &\mid Z_0,Z_1,\dots,Z_{t-1}) = Z_{t-1} + \avg(Y_{j,t}\mid Z_0,Z_1,\dots,Z_{t-1}) \\
&+\avg\Big(\avg\Big(
\sum_{l=t+1}^{k} Y_{j,l} \mid \cF_t \Big)\mid Z_0,\dots,Z_{t-1}\Big)\\
&\hspace*{.2in}-\avg\Big( \avg \Big( \sum_{l=t}^{k} Y_{j,l} \mid
\cF_{t-1} \Big)\mid Z_0,\dots,Z_{t-1}\Big)\\
& = Z_{t-1} +
\avg\Big(Y_{j,t} \mid Z_0,\dots,Z_{t-1} \Big)\\
&\hspace*{.2in}+\avg\Big( \sum_{l=t+1}^{k} Y_{j,l} \mid Z_0,\dots,Z_{t-1}\Big) \\
&\hspace*{.2in}-\avg
\Big( \sum_{l=t}^{k} Y_{j,l} \mid Z_0,\dots,Z_{t-1} \Big)\\
&= Z_{t-1},
\end{align*}
which is what we claimed.

Next we prove a bound on the random variable $|Z_t-Z_{t-1}|$.  We have
  \begin{align*}
  |Z_t-Z_{t-1}|  &= \Big|\avg \Big( \sum_{l=1}^{k} Y_{j,l} \mid \cF_t
\Big) - \avg \Big( \sum_{l=1}^{k} Y_{j,l} \mid \cF_{t-1}
\Big)\Big|\\ 
&\le \max_{a,b} \Big|\avg \Big( \sum_{l=1}^{k} Y_{j,l} \mid
\cF_{t-1}, Y_{t,l}=a \Big)
\\ &\hspace*{.4in}- \avg \Big( \sum_{l=1}^{k} Y_{j,l} \mid
\cF_{t-1}, Y_{t,l} =b \Big)\Big|\\ 
&= \max_{a,b}
\Big|\sum_{l=1}^{k}\Big(\avg \Big( Y_{j,l} \mid \cF_{t-1}, Y_{t,l}=a \Big) \\
&\hspace*{.4in}-
\avg \Big( Y_{j,l} \mid \cF_{t-1}, Y_{t,l} =b \Big)\Big)\Big|\\
&
= \max_{a,b}
\Big| a-b + \sum_{l=t+1}^{k}\Big(\avg \Big( Y_{j,l} \mid \cF_{t-1},
Y_{t,l}=a \Big) \\
&\hspace*{.4in}- \avg \Big( Y_{j,l} \mid \cF_{t-1}, Y_{t,l} =b
\Big)\Big)\Big|\\
&\le \Big| 2\mu^2 
+ \sum_{l=t+1}^{k}  \frac{2\mu^2}{N-l-2}\Big|\\
&= 2\mu^2 \frac{N-2}{N-k-2}.\end{align*}
\end{IEEEproof}

\begin{proposition} \label{prop:AH}{\rm (Azuma-Hoeffding, e.g., \cite{mcd89})} Let $X_0,\dots,X_{k-1}$ be a martingale
with $|X_{i}-X_{i-1}|\le a_i$ for each $i$, for suitable constants $a_i.$
Then for any $\nu>0,$
  $$
  \Pr\Big(\Big|\sum_{t=1}^{k-1} (X_i-X_{i-1})\Big|\ge \nu\Big)\le 2\exp \frac{-\nu^2}{2\sum a_i^2}.
  $$
\end{proposition}

{\em Proof of Theorem \ref{thm:sinc}:} Bounding large deviations for the 
sum $|\sum_{t=1}^{k}(Z_t-Z_{t-1})|=|Z_{k}-Z_0|,$
 we obtain
   \begin{equation}\label{eq:num}
   \Pr(|Z_{k} -Z_0|>  \nu)
        \le 2\exp\Big(-\frac{\nu^2}{8\mu^4k(\frac{N-2}{N-k-2})^2} \Big),
   \end{equation}
   where the probability is computed with respect to the choice of {\em ordered} $(k+1)$-tuples in $[N]$ and $\nu>0$ is any constant. 
   %Assume coherence invariance.
Using \eqref{eq:ce} and the inequality $(N-2)/(N-k-2)<2$ valid for all $k<\frac N2-1,$ we obtain
  $$
  \Pr(Z_k\ge \nu+k\bar\mu^2)\le\Pr(|Z_k-k\bar\mu^2|\ge \nu)\le 2\exp \Big(-\frac{\nu^2}{32\mu^2 k}\Big).
  $$
Now take $\beta>0$ and $\nu=\frac\beta{\log(2N/\epsilon)}-k\bar\mu^2.$
Suppose that for some $a\in(0,1)$
   \begin{equation}\label{eq:e1}
   k\mu^4 \le \frac{((1-a)\beta)^2}{32}\Big(\log\frac{2N}{\epsilon}\Big)^{-3} \quad\text{and}\quad
      k\bar\mu^2\le \frac{a\beta}{\log(2N/\epsilon)}, 
   \end{equation}
then we obtain
    \begin{equation}\label{eq:e2}
  \Pr \Big(\|\Phi_{I_j}^T\phi_j\|_2^2\ge \frac\beta{\log(2N/\epsilon)}\Big)\le  2\exp \Big(-\frac{\nu^4}{32\mu^4 k}\Big)
  \le \frac\epsilon N
  \end{equation}
Now the first claim of Theorem \ref{thm:sinc} follows by the union bound with respect to the choice of the
index $j$. 

The above proof contains the following statement.\begin{corollary}\label{cor:some}
Let $\Phi$ be an $m\times N$ matrix with mutual coherence $\mu$ and mean square coherence $\bar\mu^2$.
Let $a\in(0,1)$ and $\beta>0$ be any constants. 
Suppose that for $\alpha<\beta \log_2e,$
  $$
    \mu^4\le \frac{(1-a)^2\alpha^3}{32\beta k},\quad k\bar{\mu}^2\le a\alpha.
  $$
  Then $P_{R_k'}(\sum_{l=1}^k\mu_{i_l,j}^2\ge\alpha)\le2e^{-\beta/\alpha}.$
\end{corollary}
\begin{IEEEproof} Denote $\alpha=\beta/(\log(2N/\epsilon)),$ then $\epsilon/N=2e^{-\beta/\alpha}.$ The claim is obtained
by substituting $\alpha$ in \eqref{eq:e1}-\eqref{eq:e2}. \end{IEEEproof}

\subsection{Proof of Theorem \ref{thm:main}} 
We are now ready to prove the main Theorem \ref{thm:main}.
% and  the rest of this section is devoted to the proof.
%\begin{theorem}\label{thm:1/4} Let $\Phi$ be an $m\times N$ matrix.
%Let $\epsilon<\min\{1/k,e^{1-1/\log 2}\}$ and suppose that $\Phi$ satisfies
%  \begin{equation}\label{eq:sc}
%  k\mu^4\le \frac 1{\log^2(1/\epsilon)}\min\Big(\frac{(1-a)^2b^2}{32\log(2k)\log(e/\epsilon)},{c^2}\Big)\quad
%\text{and}\quad k\bar{\mu}^2\le\frac{ab}{\log(1/\epsilon)},
%       \end{equation}
%where $a,b,c\in(0,1)$ are constants such that
%  \begin{equation}\label{eq:abc}
%  \sqrt{a}+\sqrt{2ab}+\sqrt c+\frac {2k}N\|\Phi\|^2\le e^{-1/4}\delta/{6\sqrt 2}.
%  \end{equation}
%Then $\Phi$ is $(k,\delta,\epsilon)$-StRIP.
%\end{theorem}
The proof relies on several results from \cite{tro08}. The following theorem is a modification
of Theorem 25 in that paper. Below $R$ denotes a linear operator that performs a restriction
to $k$ coordinates chosen according to some rule (e.g., randomly). Its domain is determined by the context.
Its adjoint $R^\ast$ acts on $\reals^k$ by padding the $k$-vector with the appropriate number of zeros.
\begin{theorem}\label{thm:dec} {\rm (Decoupling of the spectral norm)} Let $A$ be a $2N\times 2N$ symmetric matrix
with zero diagonal. Let $\eta\in\{0,1\}^{2N}$ be a random vector with $N$ components equal to one.
Define the index sets $T_1(\eta)=\{i:\eta_i=0\}, T_2(\eta)=\{i:\eta_i=1\}.$ Let $R$ be a random restriction to $k$ coordinates.
For any $q\ge 1$ we have
  \begin{equation}\label{eq:RR}
    (\avg\|RAR^\ast\|^q)^{1/q}\le 2\max_{k_1+k_2=k}\avg_{\eta}(\avg\|R_1A_{T_1(\eta)\times T_2(\eta)}R_2^\ast\|^q)
^{1/q},
  \end{equation}
where $A_{T_1(\eta)\times T_2(\eta)}$ denotes the submatrix of $A$ indexed by $T_1(\eta)\times T_2(\eta)$
and the matrices $R_i$ are independent restrictions to $k_i$ coordinates from $T_i,i=1,2.$

When $A$ has order $(2N+1)\times(2N+1),$ then an analogous result holds for partitions into blocks of
size $N$ and $N+1.$
\end{theorem}
 Inequality \eqref{eq:RR} {appeared} in the proof of the decoupling theorem, Theorem 9 in 
 \cite{tro08}. The ideas behind it are due to \cite{led91}.

The next lemma is due to Tropp \cite{tro08b} and Rudelson and Vershinin \cite{rud07}.
\begin{lemma}\label{lemma:q} Suppose that $A$ is a matrix with $N$ columns and let
$R$ be a random restriction to $k$ coordinates. Let $q\ge 2, p=\max(2,2\log(\rank AR^\ast),q/2).$ Then
   $$
   (\avg\|AR^\ast\|^q)^{1/q}\le 3\sqrt p(E\|AR^\ast\|^q_{1\to 2})^{1/q}+\sqrt{\frac kN}\|A\|
   $$
   where $\|\cdot\|_{1\to 2}$ is the maximum column norm.
\end{lemma}
{The following lemma is a simple generalization of Proposition 10 in \cite{tro08}.  The only difference is that we allow the  $\xi_q$ below to be a function of $q$ instead of a constant}.
\begin{lemma}\label{lemma:LD} Let $q,\lambda>0$ and let $\xi_q$ be a positive function of $q$.
Suppose that $Z$ is a positive random variable whose $q$th moment satisfies the bound
  $$
  (\avg Z^q)^{1/q}\le \xi_q \sqrt q+\lambda.
  $$
Then
  $$
  P(Z\ge e^{1/4}(\xi_q \sqrt q +\lambda))\le e^{-q/4}.
  $$
\end{lemma}
\begin{IEEEproof}
By the Markov inequality,
  \begin{align*}
  P\left(Z\ge e^{1/4}(\xi_q \sqrt q+\lambda)\right)&\leq \frac{\avg Z^q}{(e^{1/4}(\xi_q\sqrt q+\lambda))^q} \\
  &\leq \left(\frac{\xi_q\sqrt{q}+\lambda}{e^{1/4}(\xi_q\sqrt q+\lambda)}\right)^q=e^{-q/4}. \hspace*{1in}
  \end{align*}
  \end{IEEEproof}

The main part of the proof is contained in the following lemma.
\begin{lemma}\label{lemma:technical}
Let $\Phi$ be an $m\times N$ matrix with mutual coherence parameter $\mu.$ Suppose that for some $0< \epsilon_1,\epsilon_2<1$
  \begin{equation}\label{eq:asp}
   P_{R_k'}(\{(I,i):\|\Phi_I^T\phi_i\|^2\ge \epsilon_1\}\mid i)\le \epsilon_2.
   \end{equation}
Let $R$ be a random restriction to $k$ coordinates and $H=\Phi^T\Phi-\text{Id}.$ For any $q\ge 2, p=\max(2,2\log(\rank RHR^\ast),q/2)$ we have
  \begin{align}\label{eq:RHR}
  (\avg\|R H R^\ast\|^q)^{1/q} &\le 6\sqrt p (\sqrt \epsilon_1 +(k \epsilon_2)^{1/q} \mu \sqrt{k}\nonumber\\
     &+\sqrt {2k\bar{\mu}^2}\,)+\frac {2k}N\|\Phi\|^2.
  \end{align}
\end{lemma}
\begin{IEEEproof} We begin with setting the stage to apply Theorem \ref{thm:dec}. Let $\eta\in\{0,1\}^{N}$ be a random vector with $N/2$ ones and 
let $R_1,R_2$ be random restrictions to $k_i$ coordinates in the sets $T_i(\eta),i=1,2$, respectively.
Denote by $\supp(R_i),i=1,2$ the set of indices selected by $R_i$ and let $H(\eta):=H_{T_1(\eta)\times T_2(\eta)}$.
Let $q\ge 1$ and let us bound the term $\avg_\eta(\avg\|R_1H(\eta)R_2\|^q)^{1/q}$ that appears on the right side of \eqref{eq:RR}.
The expectation in the $q$-norm is computed for two random restrictions $R_1$ and $R_2$ that are conditionally independent
given $\eta.$ Let $\avg_i$ be the expectation with respect to $R_i,i=1,2$. Given $\eta$ we can evaluate these
expectations in succession and apply Lemma \ref{lemma:q} to $\avg_2:$
  \begin{align*}
  \avg_\eta(\avg \|R_1H(\eta)R_2^\ast\|^q)^{1/q}&=
  \avg_\eta\Big[\avg_1(\avg_2\|R_1H(\eta)R_2^\ast\|^q)^{q/q}\Big]^{1/q}\\
 &\le \avg_\eta\Big\{\avg_1\Big[ 3\sqrt p\, (\avg_2\|R_1 H(\eta)R_2^\ast\|_{1\to 2}^q)^{1/q} \\
 &\hspace{0.4in}+\sqrt{\frac {2k_2}{N} }
 \|R_1H(\eta)\|\Big]^q\Big\}^{1/q}\\
 &\le \avg_\eta\Big\{3\sqrt p \;
 \Big[\avg_1\big(\avg_2\|R_1 H(\eta)R_2^\ast\|_{1\to 2}^q)\Big]^{1/q}\\
 &\hspace{0.4in}+ \sqrt{\frac {2k_2}{N} }
 \Big[\avg_1\|R_1H(\eta)\|^q\Big]^{1/q}\Big\}
  \end{align*}
where on the last line we used the Minkowski inequality (recall that the random variables involved are finite). 
Now use Lemma \ref{lemma:q} again to obtain
 \begin{align} 
   \avg_\eta(\avg \|R_1H(\eta)R_2^\ast\|^q)^{1/q}&\le 3\sqrt p\, \avg_\eta\Big[\avg_1\avg_2
   \|R_1H(\eta)R_2^\ast\|_{1\to 2}^q\Big]^{1/q}\nonumber\\
  & +3\sqrt{\frac{2k_2p}N}\avg_\eta\big(\avg_1\| H(\eta)^\ast R_1^\ast\|_{1\to2}^q\big)^{1/q}\nonumber\\
   &+\sqrt{\frac{4k_1k_2}{N^2}}\avg_\eta\|H(\eta)^\ast\|.\label{eq:terms}
  \end{align}
  Let us examine the three terms on the right-hand side of the last expression.
Let $\eta(R_2)$ be the random vector conditional on the choice of $k_2$ coordinates. The sample
space for $\eta(R_2)$ is formed of all the vectors $\eta\in\{0,1\}^{N}$ such that $\supp(R_2)\subset T_2(\eta).$
In other words, this is a subset of the sample space $\{0,1\}^N$ that is compatible with a given $R_2.$
The random restriction $R_1$ is still chosen out of $T_1(\eta)$ independently of $R_2.$
Denote by $\tilde R$ a random restriction to $k_1$ indices in the set $(\supp(R_2))^c$ and let $\tilde\avg$
be the expectation computed with respect to it. We can write
  \begin{align*}
   \avg_\eta(\avg_1\avg_2\|R_1H(\eta)&R_2^\ast\|_{1\to 2}^q)^{1/q}\\
    &\le (\avg_\eta\avg_1\avg_2\|R_1H(\eta)R_2^\ast\|_{1\to 2}^q)^{1/q}\\
       & =(\avg_2\tilde\avg
   \|\tilde R H(\eta)R_2^\ast\|_{1\to 2}^q)^{1/q}.
   \end{align*}
Recall that $H_{ij}=\mu_{ij}{\mathbbm 1}_{\{i\ne j\}}$ and that $\tilde R$ and $R_2$ are $0$-$1$ matrices.
Using this in the last equation, we obtain
  \begin{equation}\label{eq:sum}
  \avg_2\tilde\avg\|\tilde RH(\eta)R_2^\ast\|_{1\to 2}^q\le \avg_2\tilde\avg
  \max_{j\in\supp(R_2)}\textstyle{\Big(\sum_{i\in\supp(\tilde R)}\mu_{ij}^2\Big)^{q/2}}.
  \end{equation}
Now let us invoke assumption \eqref{eq:asp}. Recalling that $k_1<k,$ we have
  $$
  P_{R_2,\tilde R}\Big( \textstyle {\max\limits_{j\in\supp(R_2)}\sum_{i\in\supp(\tilde R)}\mu_{ij}^2}\ge \epsilon_1\Big)
  \le k_2\epsilon_2.
  $$
Thus with probability $1-k_2\epsilon_2$ the sum in \eqref{eq:sum} is bounded above by $\epsilon_1.$ For the
other instances we use the trivial bound $k_1\mu^2.$ We obtain
  \begin{align*}
 3\sqrt p\, \avg_\eta&\avg_1(\avg_2\|R_1H(\eta)R_2^\ast\|_{1\to 2}^q)^{1/q}\\
 &\le 3\sqrt p ((1-k_2\epsilon_2)\epsilon_1^{q/2}
 +k_2\epsilon_2(k_1\mu^2)^{q/2})^{1/q}\\
 &\le 3\sqrt p (\epsilon_1^{q/2}+k_2\epsilon_2(k_1\mu^2)^{q/2})^{1/q}\\
 &\le 3\sqrt p (\sqrt{\epsilon_1}+(k\epsilon_2)^{1/q} \sqrt{k_1\mu^2}),
 \end{align*} 
where in the last step we used the inequality $a^q+b^q
\le (a+b)^q$ valid for all $q\ge 1$ and positive $a,b.$
Let us turn to the second term on the right-hand side of \eqref{eq:terms}. We observe that
  \begin{align*}
  \|H(\eta)^\ast R_1^\ast\|_{1\to2}&=\max_{j\in T_1(\eta)}\|H_{j,T_2(\eta)}\|_2\\
  &\le \max_{j\in[N]}\|H_{j,\cdot}\|_2=\sqrt{N\bar\mu^2}
  \end{align*}
  where $H_{j,\cdot}$ denotes the $j$th row of $H$ and $H_{j,T_2(\eta)}$ is a restriction of the $j$th row to the 
  indices in $T_2(\eta).$ 
  
Finally, the third term in \eqref{eq:terms} can be bounded as follows:
    \begin{align*}
    \sqrt{\frac{4k_1k_2}{N^2}}\avg_\eta\|H(\eta)\|&\le \sqrt{\frac{(k_1+k_2)^2}{N^2}} \|H\|=\frac kN \|\Phi^T\Phi-I_N\|\\
    &\le \frac kN\max(1,\|\Phi\|^2-1)\le \frac kN\|\Phi\|^2,
    \end{align*}
    where the last step uses the fact that the columns of $\Phi$ have unit norm, and so
    $\Phi^2\ge N/m>1.$
    
    Combining all the information accumulated up to this point in \eqref{eq:terms}, we obtain
    \begin{align*}
   \avg_\eta&(\avg \|R_1H(\eta)R_2^\ast\|^q)^{1/q}\\
   &\hspace{0.3in}\le 3\sqrt p(\sqrt{\epsilon_1}+(k\epsilon_2)^{1/q}\mu\sqrt k+
   \sqrt{2k_2\bar{\mu}^2}\,)+\frac kN\|\Phi\|^2.
 \end{align*}
   Finally, use this estimate in \eqref{eq:RR} to obtain the claim of the lemma.
\end{IEEEproof}

\begin{IEEEproof}[Proof of Theorem \ref{thm:main}]
The strategy is to fix a triple $a,b,c\in (0,1)$ that satisfies \eqref{eq:abc} and to prove that \eqref{eq:sc} 
implies $(k,\delta,\epsilon)$-StRIP. 
Let $\epsilon_1=\frac{b}{\log1/\epsilon}$ and $\epsilon_2=k^{-1+\log\epsilon}$. In Corollary \ref{cor:some} set 
$\alpha=\epsilon_1$ and $\beta=\alpha\log(2/\epsilon_2).$  Under the assumptions in \eqref{eq:sc} 
this corollary implies that
   $$
    P_{R'}\Big(\sum\limits_{m=1}^k \mu_{i_m,j}^2>\epsilon_1\Big)<\epsilon_2.
  $$
Invoking Lemma \ref{lemma:technical}, we conclude that \eqref{eq:RHR} holds with the current values of $\epsilon_1,\epsilon_2$.
For any $q\geq 4\log k$ we have $p=q/2$, and thus \eqref{eq:RHR} becomes
 \begin{align}\label{eq:ld}
  (\avg\|R H R^\ast\|^q)^{1/q}&\le {3}\sqrt {2q} (\sqrt \epsilon_1 +(k \epsilon_2)^{1/q} \mu \sqrt{k}\nonumber\\
    & +\sqrt {2k\bar{\mu}^2})+2\frac kN\|\Phi\|^2.
   \end{align}
Introduce the following quantities:       
      $$\xi_q=3\sqrt{2}(\sqrt {\epsilon_1}+(k \epsilon_2)^{1/q} \mu \sqrt{k}
     +\sqrt {2k\bar{\mu}^2}) \ \  \text{and} \ \  \lambda=\frac {2k}N\|\Phi\|^2.
   $$   
Now \eqref{eq:ld} matches the assumption of Lemma \ref{lemma:LD}, and we obtain
    \begin{equation}\label{eq:P_RHR}
      P_{R_k}(\|R H R^\ast\| \geq e^{1/4}(\xi_q\sqrt q +\lambda))\leq e^{-q/4}.
    \end{equation}
Choose $q=4\log(1/\epsilon),$ which is consistent with our earlier assumptions on $k,q,$ and $\epsilon.$
With this, we obtain
      $$P_{R_k}\big(\|R H R^\ast\|\geq e^{1/4}(\xi_q\sqrt q +\lambda)\big)\leq \epsilon.$$
Now observe that $\|R H R^\ast\|\leq \delta$ is precisely the RIP property for the support identified
by the matrix $R.$ Let us verify that the inequality 
   \begin{align*}
    6\sqrt{2} \big(\sqrt \epsilon_1&+(k\epsilon_2)^{1/q}\sqrt{k\mu^2}\\
    &+\sqrt{2k\bar{\mu}^2}\big)\sqrt{\log(1/\epsilon)}+\frac{2k}{N}\|\Phi\|^2<e^{-1/4}\delta
    \end{align*}
   is equivalent to \eqref{eq:abc}. This is shown by substituting $\epsilon_1$ and $\epsilon_2$ with their definitions, 
   and $\mu$ and $\bar{\mu}^2$ with their bounds in statement of the theorem.
Thus, $P_{R_k}(\|R H R^\ast\|\geq \delta)\le\epsilon,$ which establishes the StRIP property of $\Phi.$
\end{IEEEproof}
\vspace*{.1in}

\remove{To see that matrices that satisfy the constraints of Theorem \ref{thm:sinc} exist,
take again the Delsarte-Goethals matrices \eqref{eq:example} with r=1.
Then $\mu=2m^{-1/2}$, so taking $k=\sqrt m=(2N)^{1/6}$ it is possible to satisfy the
constraints on $\mu$ and $m$ in \eqref{eq:constraints}.
In the next section we will see that it is possible to construct a broad class of
sampling matrices whose
parameters satisfy the assumptions of both Theorems \ref{thm:strip} and \ref{thm:sinc}.}

%%%%%%%%
%% EXAMPLES
%%%%%%%
\section{Examples and extensions}\label{sec:determin}
\subsection{Examples of sampling matrices.} It is known \cite{don09} that experimental performance of many known RIP sampling matrices in sparse recovery is far better than
predicted by the theoretical estimates. Theorems \ref{thm:sinc} and \ref{thm:main} provide some insight into the reasons for such behavior. 
As an example, take binary matrices constructed from the Delsarte-Goethals codes mentioned previously.
%\cite[p.461]{mac91}.
%The parameters of the matrices are as follows:
%  \begin{equation}\label{eq:example}
%  m=2^{2s+2}, \;N=2^{-r}m^{r+2},\;\mu=2^rm^{-\half}
%  \end{equation}
%where $s\ge 0$ is any integer, and where for a fixed $s$, the parameter $r$ can be any number
%in $\{0,1,\dots,s-1\}.$
The sampling matrices $\Phi$ obtained from them are coherence-invariant.
If we take $s$ to be an odd integer and set $r=(s+1)/2$, then we obtain for this family of matrices the parameters
   $$
m= 2^{4r},\; N=2^{4r^2+7r},\; \mu=m^{-1/4}.
   $$
%The matrix $\Phi$ is coherence-invariant, so we put $\theta=\bar\mu^2.$
%Lemma \ref{lem:pless} below implies that
%From Pless identities  (e.g. \cite[p.132]{mac91}), we have that,
%   \begin{equation}\label{eq:mu2}
%\bar\mu^2=\frac{N-m}{m(N-1)}<\frac{1}{m},
%   \end{equation}
As noted above, we have $\bar\mu^2<1/m$ and $\|\Phi\|=\sqrt{N/m}$.
Thus for $\mu$ and $\bar\mu^2$ to satisfy the assumptions in 
Theorems \ref{thm:sinc} and \ref{thm:main}, we need $m$, $ N$, and $k$ 
to satisfy the relation $m=\Theta(k \log^3 \frac{N}{\epsilon})$ which is nearly optimal for sparse-recovery.
Note that to satisfy just the assumptions of Thm.~\ref{thm:main}, we can construct a Delsarte-Goethals matrix with shorter column length of $m = O(k \log k),$ see Section~\ref{sec:dels}.

Similar logic leads to derivations of such relations for other matrices. We summarize  
these arguments in the next proposition, which shows that matrices with nearly optimal
sketch length support high-probability recovery of sparse signals chosen from the 
generic signal model (more on sparse recovery in the Appendix; see in particular Theorem \ref{theorem:bp}). 
\begin{definition}\label{def:generic}
We say that a signal $\bfx\in \reals^N$ is drawn from a {\sl generic random
signal model} $\cS_k$ if

1) The locations of the $k$ coordinates of $\bfx$ with largest magnitudes
are chosen among all $k$-subsets $I\subset [N]$ with a uniform distribution;

2) Conditional on $I$, the signs of the coordinates $x_i, i\in I$ are
i.i.d. uniform Bernoulli random variables taking values in the set $\{1,-1\}$.
\end{definition}
\begin{proposition}\label{prop:final}
Let $\Phi$ be an $m\times N$ sampling matrix.  Suppose that it has coherence parameters
$\mu=O(m^{-1/4})$, $\bar\mu^2=O(m^{-1}),$ %where $\theta=\bar\mu^2$  %or $\theta=\bar\mu_{\max}^2$
%according as $\Phi$ is coherence-invariant or not, 
and
  $$
\|\Phi\|=O(\sqrt{N/k}).
  $$
 If $m=\Theta(k(\log (N/\epsilon))^3)$ and $k<1/\epsilon,$ 
then $\Phi$ supports sparse recovery under
Basis Pursuit for all but an $\epsilon$ proportion of $k$-sparse signals chosen from the generic
random signal model $\cS_k.$
\end{proposition}
We remark that the conditions on mean square coherence are generally easy to achieve. 
As seen from Table \ref{table} below, they are satisfied by most examples considered in the existing
literature, including both random and deterministic constructions. 
 The most problematic quantity is the mutual coherence parameter $\mu$. 
It might either be large itself, or have a large theoretical bound. 
Compared to earlier work, our results rely on a more relaxed condition on $\mu$, 
enabling us to establish near-optimality for new classes of matrices.  
For readers' convenience, we summarize in Table 1 a list of such optimal matrices 
along with several of their useful properties. A systematic description of all but the last 
two classes of matrices can be found in \cite{baj11}. Therefore 
we limit ourselves to giving definitions and performing some not immediately obvious calculations 
of the newly defined parameter, the mean square coherence.

\vspace*{.1in}\emph{Normalized Gaussian Frames.} A normalized Gaussian frame is obtained
by normalizing each column of a Gaussian matrix with independent, Gaussian-distributed entries 
that have zero mean and unit variance. 
The mutual coherence and spectral norm of such matrices were characterized in \cite{baj11} (see Table \ref{table}). 
These results together with the relation $\bar\mu^2<\mu^2$ lead to a trivial upper bound on 
$\bar\mu^2$, namely $\bar\mu^2\leq 15\log N/m$. 
Since this bound is already tight enough for $\bar\mu^2$ to satisfy the assumption 
of Proposition \ref{prop:final}, and to avoid distraction from the main goals of the paper, 
we made no attempt to refine it here. 

\vspace*{.1in}\emph{Random Harmonic Frames}: Let $\mathcal{F}$ be an $N\times N$ discrete Fourier transform matrix, i.e., 
$\mathcal{F}_{j,k}=\frac{1}{\sqrt{N}}e^{2\pi i jk/N}$. Let $\eta_i$, $i=1,...,N$, be a sequence of independent Bernoulli random variables with mean $\frac{m}{N}$. Set $\mathcal{M}=\{i: \eta_i=1\}$ and use $\mathcal{F}_{\mathcal{M}}$ to denote the submatrix of $\mathcal{F}$ whose row indices lies in $\mathcal{M}$. Then the random matrix {$\sqrt{\frac{N}{|\mathcal{M}|}}\mathcal{F}_{\mathcal{M}}$ }  is called a random harmonic frame
\cite{can06a,can06b}. In the
next proposition we compute the mean square coherence for all realizations of this matrix.
\begin{proposition} All instances of the random harmonic frames are coherence invariant with the following mean square coherence
\[
\bar\mu^2 =\frac{N-|\mathcal{M}|}{(N-1)|\mathcal{M}|}.
\]
\end{proposition}
\begin{IEEEproof}
For each $t\in [|\mathcal{M}|]$, let $a_t$ with  be the $t$-th member of $\mathcal{M}$.
To prove coherence invariance, we only need to show that 
$\{\mu_{j,k}: k\in [N]\backslash j\}=\{\mu_{N,k}: k\in [N-1]\}$ holds for all $j\in [ N ]$. This is true since
\[
\mu_{j,k}=\frac{1}{|\mathcal{M}|}\sum\limits_{t=1}^{|\mathcal{M}|} e^{\frac{2\pi i (j-k) a_t}{N}}=\mu_{N,(k-j+N)\text{mod } N}
\quad \text{for all } k\neq j.
\]
In words, the $k$th coherence in the set $\{\mu_{j,k}, k\in [N]\backslash j\}$ is exactly the $\left(k-j+N\mod  N\right)$-th coherence in $\{\mu_{N,k}, k\in [N-1]\}$, therefore the two sets are equal. We proceed to calculate the mean square coherence,
\begin{align*}
\bar\mu^2&=\frac{1}{N(N-1)|\mathcal{M}|^2}\sum\limits_{j\neq k,j,k=1}^N\left|\sum\limits_{t=1}^{|\mathcal{M}|} e^{2\pi i (j-k)a_t/N}\right|^2 \\
&=\frac{1}{N(N-1)|\mathcal{M}|^2}\sum\limits_{j\neq k, j,k=1}^N \sum\limits_{t_1,t_2=1}^{|\mathcal{M}|} e^{2\pi i(j-k)(a_{t_1}-a_{t_2})/N}\\
&=\frac{1}{N(N-1)|\mathcal{M}|^2}\Big(\sum\limits_{j\neq k,j,k=1}^N \sum \limits_{t_1=t_2=1}^{|\mathcal{M}|} 1\\
&+\sum\limits_{t_1\neq t_2 ,t_1,t_2=1}^{|\mathcal{M}|}\sum\limits_{k=1}^N\sum\limits_{j\neq k} e^{2\pi i(j-k)(a_{t_1}-a_{t_2})/N}\Big)\\
&=\frac{1}{N(N-1)|\mathcal{M}|^2}(N(N-1)|\mathcal{M}|-|\mathcal{M}|(|\mathcal{M}|-1)N)\\
&=\frac{N-|\mathcal{M}|}{(N-1)|\mathcal{M}|}.
\end{align*}
\end{IEEEproof}

\vspace*{.1in}\emph{Chirp Matrices}: Let $m$ be a prime. An $m\times m^2$ ``chirp matrix'' $\Phi$ is defined by $\Phi_{t,am+b}=\frac{1}{\sqrt{m}}e^{2\pi i (bt^2+at)/m}$ for $t,a,b=1,...,m$. The coherence between each pairs of column vectors is known to be 
    $$
\mu_{jk}=\frac{1}{\sqrt{m}}     \quad(j\ne k),
   $$
 from which we immediately obtain the inequalities $\mu \leq 1/\sqrt{m}$ and $\bar\mu^2 \leq 1/m$. More details on these
 frames are given, e.g., in \cite{Brodzik06,Casazza06}.

\vspace*{.1in}\emph{Equiangular tight frames (ETFs)}: A matrix $\Phi$ is called an ETF if its 
columns $\{\phi_i \in \mathbb{R}^m, i=1,...,N\}$ satisfy the following two conditions:
\begin{itemize}
\item $\|\phi_i\|_2=1$, for $i=1,...,N$.
\item $\mu_{ij}=\sqrt{\frac{N-m}{m(N-1)}}$, for $i\neq j$.
\end{itemize}
From this definition we obtain $\mu=\sqrt{\frac{N-m}{m(N-1)}}$ and $\theta=\bar\mu^2=\frac{N-m}{m(N-1)}$.
The entry in the table also covers the recent construction of ETFs from Steiner systems \cite{Fickus12}.

\vspace*{.1in}\emph{Reed-Muller matrices:} In Table \ref{table} we list two tight frames
obtained from binary codes. The Reed-Muller matrices are obtained from certain special subcodes
of the second-order Reed-Muller codes \cite{mac91}; their coherence parameter $\mu$ is found in \cite{baj11}
and the mean square coherence is found from \eqref{eq:mu21}. The Delsarte-Goethals matrices
are also based on some subcodes of the second order Reed-Muller codes and were discussed earlier in 
this section. 
Both dictionaries %support orthogonal arrays, and therefore, 
form unit-norm tight frames (the rows
of the matrix $\Phi$ are pairwise orthogonal), with a consequence that $\|\Phi\|=\sqrt{N/m}.$
We include these two examples out of many other possibilities based on codes because they
appear in earlier works, and because their parameters are in the range 
that fits well our conditions.

We note that the quaternary version of these frames is also of interest in the context
of sparse recovery; see in particular \cite{cal10a}.

\vspace*{.05in}
\emph{Deterministic sub-Fourier Construction \cite{hau10}}: Let $p>2$ be a prime, 
and let $f(x)\in \ff_p[x]$ be a polynomial of degree $d>2$ over the finite field $\ff_p$. 
Suppose that $m$ is some integer satisfying 
$p^{1/(d-1)}\leq m\leq p$. 
Then we can construct an $m\times p$ deterministic RIP matrix from a $p\times p$ DFT matrix 
by keeping only the rows with indices in $\{f(n)\!\! \pmod p, n=1,\dots,m\},$ 
and normalizing the columns of the resulting matrix. These submatrices form tight frames, and
so their spectral norms can be easily verified to be $\sqrt{p/m}$.
It is known \cite{hau10} that this matrix has mutual coherence no greater than $e^{3d}m^{-1/(9d^2\log d)}$. 
Even though this bound is an artifact of the proof technique used in \cite{hau10}, 
there seem to be no obvious ways of improving it. 

\vspace*{.1in}{\footnotesize

\begin{table*}[t]\begin{center}
     \begin{tabular}{| l | c   c c c |}
     \hline
Name & $\mathbb{R}$/$\mathbb{C}$   & Dimensions & $\mu$  & \red{$\bar \mu^2$}   \\[.05in]
   \hline
     Normalized Gaussian (G) & $\mathbb{R}$   & $m\times N$ & $\leq \frac{\sqrt{15\log N}}{\sqrt{m}-\sqrt{12\log N}}$ &  $\leq \mu^2$
        \\[.1in]
     Random harmonic  (RH)   & $\mathbb{C}$  & $|\mathcal{M}|\times N$, $\frac{1}{2}m\leq |\mathcal{M}|\leq \frac{3}{2}m$   & $\leq \sqrt{\frac{118(N-m)\log N}{mN}}$  &      $\leq\frac{N-|\mathcal{M}|}{|\mathcal{M}|(N-1)}$ \\[.1in]
     Chirp  (C)     & $\mathbb{C}$  &  $m\times m^2$ & $\frac{1}{\sqrt m}$      &      $\frac{1}{m+1} $   \\[.05in]
     ETF (including Steiner) &  $\mathbb{C}$       &  $\sqrt{N}\leq m\leq N$  & $\sqrt{\frac{N-m}{m(N-1)}}$ &  $\mu^2$  \\[.05in]
     Reed-Muller (RM) & $\mathbb{R}$        &  $2^s\times 2^{t(1+s)}$ & $\leq \frac{1}{\sqrt{2^{s-2t-1}}}$ ,    &  $\leq 2^{-s} $      \\[.05in]
  Delsarte-Goethals set (DG)    &$\mathbb{R} $      &  $2^{2s+2}\times 2^{2(s+1)(r+2)-r}$    &$2^{r-s-1} $ &     $ \leq 2^{-2s-2}$   \\[.05in]
   Deterministic subFourier (SF) & $\mathbb{C}$     & $ m\times p$ & $e^{3d}m^{-1/(9d^2\log d)}$        & $\leq \frac{1}{m}$   \\[.05in]
   \hline
     \end{tabular}
\end{center}
\vspace*{.1in}
\begin{center}
   \begin{tabular}{| l |c c c c  |}
   \hline
    Name &      $\|\Phi\|$  &Restrictions &  Probability       & Requirement for StRIP: $m= O(\cdot)$     \\[.05in]
   \hline
    G   & $\leq \frac{\sqrt{m}+\sqrt{ N}+\sqrt{2\log N}}{\sqrt{m-\sqrt{8m\log N}}}$    &  
            $60\log N\leq m\leq \frac{N-1}{4\log N}$      &      $\geq 1-\frac{11}{N}$ &$ \max\{k,\sqrt{k\log k}\log N$\} \\[.1in]
    RH       & {$\leq \sqrt{\frac{N}{m}}$ }  & $16\log N\leq m\leq \frac{N}{3}$ & $\geq 1-\frac{4}{N}-\frac{1}{N^2}$  &$\max\{k,\sqrt{k\log k}\log N\}$  \\[.05in]    
   C & $\sqrt m$  &  $m$ is prime &      deterministic &$ k$  \\[.05in]
    ETF &  $\sqrt{\frac{N}{m}}$ & $\sqrt{\frac{M(N-1)}{N-M}}$,$\sqrt{\frac{(N-m)(N-1)}{m}}$ are odd integers  &  deterministic  &$ k $ \\[.05in] 
    RM & $2^{ts/2}$ & $t<s/4$        &     deterministic &$k$ \\[.05in]
   DG        & $2^{(s+1)(r+1)-r/2}$       & $r<s/2$     & deterministic  &$k$  \\[.05in]
  SF  &$\sqrt{p/m}$  & $p$ is prime, $p^{1/(d-1)}\leq m\leq p$      &  deterministic  &$\max\{k, (k\log k)^{\frac{9d^2\log d}{4}}\}$  \\[.05in] 
   \hline
    \end{tabular}
\end{center}\caption{Examples for Theorem.~\ref{thm:main}: Classes of sampling matrices satisfying the  StRIP.  
}\label{table}
\end{table*}}

\providecommand{\href}[2]{#2}

%\bibliography{sensing}
%\bibliographystyle{amsplain}

%\begin{appendices}
%\section{Appendix}
\appendix

\section{Error bounds for recovery by
Basis Pursuit}\label{sec:app}
Among the most studied estimators for sparse recovery  is the Basis Pursuit algorithm
\cite{che98}.
This is an $\ell_1$-minimization algorithm that provides an
estimate of the signal through solving a convex programming problem
  \begin{equation}\label{eq:bp}
   \hat\bfx=\arg\min\|\tilde\bfx\|_{1} \text{\quad subject to }
   \Phi\tilde\bfx=\bfy.
  \end{equation}
In this section we prove approximation error bounds for recovery by
Basis Pursuit from linear sketches obtained using
deterministic matrices with the StRIP and SINC properties.

%\subsection{StRIP Matrices with incoherence property} 
It was proved in \cite{tro08} that random 
sparse signals sampled using matrices with the StRIP property can be 
recovered with high probability from low-dimensional sketches using linear programming. 
\red{ Theorem \ref{theorem:bp} below generalizes this result to signals that are not necessarily
sparse.
Its proof essentially follows from \cite{can06a} with an extra calculation of the failure
rate stemming from replacing the hard RIP condition with its statistical version. It is presented here
for reader's convenience.}

\begin{theorem}\label{theorem:bp}
Suppose that $\bfx$ is a generic random signal from the model
$\cS_k.$ Let
$\bfy=\Phi\bfx$ and let $\hat\bfx$ be the approximation of $\bfx$ by the Basis
Pursuit algorithm.
Let $I$ be the set of $k$ largest coordinates of $\bfx$. If
  \begin{enumerate}
  \item $\Phi$ is $(k,\delta,\epsilon)$-StRIP;
  \item $\Phi$ is $(k,\frac{(1-\delta)^2}{8\log(2N/\epsilon)},\epsilon)$-SINC,
\end{enumerate}
then with probability at least $1-3\epsilon$
  \begin{equation}\label{eq:sup}
  \|\bfx_I-\hat\bfx_I\|_{2}\le \frac{1}{2\sqrt{2\log(2N/\epsilon)}}
      \min_{\bfx'\text{\rm is $k$
  -sparse} }\|\bfx- \bfx' \|_{1}
  \end{equation}
and
  \begin{equation}\label{eq:offsup}
\|\bfx_{I^c}-\hat{\bfx}_{I^c}\|_{1} \le 4 \min_{\bfx' \text{\rm is $k$
  -sparse} }\|\bfx- \bfx' \|_{1}
  \end{equation}
\end{theorem}
This theorem implies that if the signal $\bfx$ itself is $k$-sparse then
the basis pursuit algorithm will recover it exactly. Otherwise, its output
$\hat \bfx$ will be a tight sparse approximation of $\bfx$.
\red{Note that it is easy to join the estimates \eqref{eq:sup} and \eqref{eq:offsup} into a single inequality
that gives an $l_2/l_1$ error guarantee.}

Theorem \ref{theorem:bp} will follow from the next three lemmas. Some of the
ideas involved in their proofs are close to the techniques used in \cite{can06a}.
Let $\bfh=\bfx-\hat\bfx$ be the error in recovery of basis pursuit.
In the following $I\subset[N]$ refers to the support of the $k$ largest coordinates
of $\bfx.$ 
\begin{lemma}\label{lemma:error-sup} Let $s=8\log(2N/\epsilon).$
Suppose that
   $\|(\Phi_I^T\Phi_I)^{-1}\|\le\frac1{1-\delta}$
and
  $$
   \|\Phi_I^T\phi_i\|_{2}^2\le s^{-1}(1-\delta)^2 \quad\text{for all }i\in I^c := [N]\setminus I.
  $$
Then
  $$
  \|\bfh_I\|_{2}\le s^{-\half}\,\|\bfh_{I^c}\|_{1}.
  $$
\end{lemma}
\begin{IEEEproof} Clearly,
  $
   \Phi\bfh=\Phi\hat\bfx-\Phi\bfx=0,
  $
so $\Phi_I\bfh_I=-\Phi_{I^c}\bfh_{I^c}$ and
  $$
  \bfh_I=-(\Phi_I^T\Phi_I)^{-1}\Phi_I^T\Phi_{I^c}\bfh_{I^c}.
  $$
We obtain
  \begin{align*}
    \|\bfh_I\|_{2}&\le \|(\Phi_I^T\Phi_I)^{-1}\|\|\Phi_I^T\Phi_{I^c}\bfh_{I^c}\|_{2}\le
   \frac1{1-\delta}\sum_{i\in I^c}\|\Phi_I^T\phi_i\|_{2}|h_i|\\
  &\le  s^{-\half}\,\|\bfh_{I^c}\|_{1},
 \end{align*}
as required.
\end{IEEEproof}

Next we show that the error outside $I$ cannot be large. Below $\sgn(\bfu)$ is a $\pm1$-vector
of signs of the argument vector $\bfu.$
\begin{lemma} \label{lemma:v}
  Suppose that there exists a vector $\bfv\in\reals^N$ such that
  \begin{enumerate}
   \item[(i)] $\bfv$ is contained in the row space of $\Phi$, say $\bfv=\Phi^T\bfw;$
   \item[(ii)] $\bfv_I=\sgn(\bfx_I);$
   \item[(iii)] $\|\bfv_{I^c}\|_\ellinf\le \half.$
  \end{enumerate}
Then
  \begin{equation}\label{eq:h}
    \|\bfh_{I^c}\|_{1}\le 4\|\bfx_{I^c}\|_{1}.
  \end{equation}
\end{lemma}
\begin{IEEEproof}
By \eqref{eq:bp} we have
  \begin{align*}
    \|\bfx\|_{1}&\ge \|\hat\bfx\|_{1}=\|\bfx+\bfh\|_{1}
   =\|\bfx_I+\bfh_I\|_{1}+\|\bfx_{I^c}+\bfh_{I^c}\|_{1}\\
    &\ge \|\bfx_I\|_{1}+\ip{\sgn(\bfx_I)}{\bfh_I}+\|\bfh_{I^c}\|_{1}-\|\bfx_{I^c}\|_{1}.
  \end{align*}
Here we have used the inequality $\|\bfa+\bfb\|_{1}\ge\|\bfa\|_{1}+\ip{\sgn(\bfa)}
{\bfb}$ valid for any two vectors $\bfa,\bfb\in\reals^N$ and the triangle inequality.
From this we obtain
  $$
  \|\bfh_{I^c}\|_{1}\le |\ip{\sgn(\bfx_I)}{\bfh_I}|+2\|\bfx_{I^c}\|_{1}.
  $$
Further, using the properties of $\bfv,$ we have
\begin{eqnarray*}
|\langle\sgn(\bfx_I),\bfh_I\rangle| &=& |\langle\bfv_I,\bfh_I\rangle|\\
&=& |\langle\bfv,\bfh\rangle - \langle\bfv_{I^c},\bfh_{I^c}\rangle|\\
&\le& |\langle \Phi^T \bfw,\bfh\rangle| + |\langle\bfv_{I^c},\bfh_{I^c}\rangle|\\
&\le& |\langle \bfw,\Phi \bfh\rangle| + \|\bfv_{I^c}\|_{\ell_\infty} \|\bfh_{I^c}\|_{1}\\
&\le& \frac12  \|\bfh_{I^c}\|_{1}.
\end{eqnarray*}
The statement of the lemma is now evident.
\end{IEEEproof}

Now we prove that such a vector $\bfv$ as defined in the last lemma indeed exists.
\begin{lemma} \label{lemma:exists_v} Let $\bfx$ be a generic random signal from the model $\cS_k.$
Suppose that the support $I$ of the $k$ largest coordinates of $\bfx$ is fixed.
Under the assumptions of Lemma \ref{lemma:error-sup} the vector
  $$
   \bfv=\Phi^T\Phi_I(\Phi_I^T\Phi_I)^{-1}\sgn(\bfx_I)
  $$
satisfies (i)-(iii) of Lemma \ref{lemma:v} with probability at least $1-\epsilon.$ 

\end{lemma}
\begin{IEEEproof}
From the definition of $\bfv$ it is clear that it belongs to the row-space of $\Phi$ and
$\bfv_I =\sgn(\bfx_I).$
We have $v_i = \phi_i^T \Phi_I (\Phi_I^T\Phi_I)^{-1}\sgn(\bfx_I) = \ip{\bfs_i}{\sgn(\bfx_I)},$
where
   $$
\bfs_i = (\Phi_I^T\Phi_I)^{-1}\Phi_I^T \phi_i \in \reals^k.
   $$
We will show that $|v_i|\le \frac12$ for all $i \in I^c$ with
probability $1-\epsilon.$

Since the coordinates of $\sgn(\bfx_I)$ are i.i.d. uniform random variables taking
values in the set $\{\pm1\}$, we can use Hoeffding's inequality to claim that
\begin{equation}\label{Hoeffding}
P_{R^k}(|v_i| >1/2 )\le 2\exp\Big(-\frac1{8\|\bfs\|_2^2}\Big).
\end{equation}
On the other hand, for all $i\in I^c,$
\begin{eqnarray}\label{magnitude}
\|\bfs_i\|_{2} &=& \|(\Phi_I^T\Phi_I)^{-1}\Phi_I^T \phi_i\|_{2}\notag\\
&\le & \|(\Phi_I^T\Phi_I)^{-1}\| \|\Phi_I^T \phi_i\|_{2}\notag\\
&\le& \frac1{1-\delta} \frac{1-\delta}{\sqrt{8\log(2N/\epsilon)}}\notag\\
&=& \frac1{\sqrt{8\log(2N/\epsilon)}}.
\end{eqnarray}
Equations \eqref{Hoeffding} and \eqref{magnitude} together imply for any $i\in I^c,$
$$
P_{R^k}\Big(|v_i| >\frac12\Big) \le 2\exp\Big(-\frac1{8(1/\sqrt{8\log(2N/\epsilon)})^2}\Big)= \frac{\epsilon}N.
$$
Using the union bound, we now obtain the following relation:
   \begin{equation}\label{eq:1/2}
 P_{R^k}\Big(\|\bfv_{I^c}\|_{\infty}>1/2\Big)\le \epsilon.
   \end{equation}
Hence $|v_i|\le \frac12$ for all $i \in I^c$ with probability at least $1-\epsilon$.
\end{IEEEproof}

Now we are ready to prove Theorem~\ref{theorem:bp}.
\vspace{.1in}

\begin{IEEEproof}[Proof of Theorem~\ref{theorem:bp}]
The matrix $\Phi$ is $(k,\delta,\epsilon)$-SRIP. Hence,
with probability at least $1-\epsilon,$ $\|(\Phi_I^T\Phi_I)^{-1}\| \le \frac{1}{1-\delta}$.
At the same time, from the SINC assumption we have, with probability at least $1-\epsilon$
over the choice of $I$,
  $$
\|\Phi_I^T\phi_i\|_{2}^2 \le \frac{(1-\delta)^2}{8\log(2N/\epsilon)},
  $$
for all $i \in I^c.$ Thus, $\Phi_I$ will have these two properties
with probability at least $1-2\epsilon$.
Then from Lemma~\ref{lemma:error-sup} we obtain that
 $$
 \|\bfh_I\|_2 \le \frac{1}{\sqrt{8\log(2N/\epsilon)}} \|\bfh_{I^c}\|_1,
 $$
with probability $\ge 1-2\epsilon.$ Furthermore, from Lemmas \ref{lemma:v}, \ref{lemma:exists_v}
$$
\|\bfh_{I^c}\|_1 \le 4 \|\bfx_{I^c}\|_1,
$$
with probability $1-\epsilon$. This completes the proof.
  \end{IEEEproof}

\begin{IEEEbiographynophoto}
{Alexander Barg} (Mâ'00â SMâ'01â Fâ'08) received the M.Sc. degree in applied
mathematics and the Ph.D. degree in electrical engineering, the latter from the Institute 
for Information Transmission Problems (IPPI) Moscow, Russia, in 1987. He has been a Senior 
Researcher at the IPPI since 1988. He spent years 1995-1996 at the Technical University of
Eindhoven, Eindhoven, the Netherlands. During 1997-2002, he was Member
of Technical Staff of Bell Labs, Lucent Technologies. Since 2003 he has been
a Professor in the Department of Electrical and Computer Engineering and
Institute for Systems Research, University of Maryland, College Park. 

Alexander Barg was the recipient of the IEEE Information Theory Society Paper Award in 2015. During 1997-2000, A. Barg was an Associate Editor for Coding Theory
of the IEEE TRANSACTIONS ON INFORMATION THEORY. He was the Technical
Program Co-Chair of the 2006 IEEE International Symposium on Information
Theory and of 2010 and 2015 IEEE ITWs. He serves on the Editorial Board 
of several journals
including Problems of Information Transmission, SIAM Journal on Discrete
Mathematics, and Advances in Mathematics of Communications. 

Alexander Barg's research interests are in coding and information theory, 
signal processing, and algebraic combinatorics.
\end{IEEEbiographynophoto}

\begin{IEEEbiographynophoto}{Arya Mazumdar} (S'05-M'13)
 is an assistant professor in University of Minnesota-Twin Cities (UMN). Before coming to UMN, he was a postdoctoral scholar
  at the Massachusetts Institute of Technology. He received the Ph.D. degree 
   from the University of Maryland, College Park, in 2011. 

Arya is a recipient of the NSF CAREER award, 2015 and  the 2010 IEEE ISIT Student Paper Award. He is also the recipient of the Distinguished Dissertation Fellowship Award, 2011, at the University of Maryland.    
   He spent the summers of 2008 and 2010 at the Hewlett-Packard Laboratories, Palo Alto, CA, and IBM Almaden Research Center, San Jose, CA, respectively.
 Arya's research interests include error-correcting codes, information theory and their applications.
 \end{IEEEbiographynophoto}
\begin{IEEEbiographynophoto}{Rongrong Wang} 
received the B.S. degree in mathematics from Peking University, China, in 2007, and the PhD degree in applied mathematics from the University of Maryland College Park in 2013. She is now a postdoctoral researcher at the University of British Columbia. Her main research interest includes compressed sensing, Sigma Delta quantization, frame theory, and seismic inverse problems.
\end{IEEEbiographynophoto}

\balance
\end{document}